\documentclass[12pt]{article}

\usepackage{amsmath,amssymb,amsfonts,bm}
\usepackage{dcolumn}
\usepackage{mathrsfs}

\usepackage{graphicx}

\begin{document}

\begin{center}
{\Large\textbf{Monte Carlo simulations of $\bm{\mu}$CF processes
 kinetics in~deuterium gas}}

\vspace*{6mm}
{Andrzej Adamczak}\\
{\small\textit{Institute of Nuclear Physics, Polish Academy of Sciences,
 Krak\'ow, Poland,\\
 and Rzesz\'ow Technical University, Rzesz\'ow, Poland,}}\\[4pt]
Mark~P.~Faifman\\
{\small\textit{Russian Research Center ``Kurchatov Institute'', 
 Moscow, Russia.}}

\end{center}

\vspace*{6mm}

\begin{abstract}
  The muon-catalyzed-fusion processes in D$_2$ gas for various
  temperatures and densities have been studied by means of
  Monte Carlo simulations. In particular, the role of the resonant
  and nonresonant $dd\mu$ formation and differences between the
  neutron time spectra from the ortho-D$_2$ and para-D$_2$ targets
  at low temperatures have been investigated.
\end{abstract}

\vspace*{6mm}

The negative muons which slow down in a~D$_2$ target form the $d\mu$
muonic hydrogen atoms in the highly excited Rydberg states. Then, such
atoms undergoing the various cascade processes~\cite{leon62} deexcite to
the ground 1S~state and participate in numerous processes accompanying
$\mu$CF (muon-catalyzed-fusion) phenomenon~\cite{mens87}. The
ground-state 1S atoms can gain high kinetic energies up to
$\sim{}1000$~eV, as a~result of several accelerating collision processes
during the atomic cascade (see, e.g.,
Refs.~\cite{abbo97,pohl01,faif01,jens02}). However, a~fraction of the
1S muonic atoms is thermalized after the cascade.

The aim of this work is to study the role of ``hot'' $d\mu$ atoms,
supplied by the cascade, and the competition between the
energy-dependent processes (muonic atom scattering and muonic molecule
formation) of $\mu$CF cycle in a~pure deuterium D$_2$ gas.

The D$_2$ targets simulated in our Monte Carlo calculations correspond
to the D$_2$ targets which are to be used at JINR in the upcoming
experiment~\cite{demi05} on the radiative capture reaction
$d+d\to{}^4\mathrm{He}+\gamma$ in the $dd\mu$ muonic molecule.  The
deceleration, thermalization and diffusion of $d\mu$'s in D$_2$ perfect
gas is described using the differential scattering cross sections. At
collision energies~$\varepsilon$ greater than about~1~eV, the doubled
cross section for $d\mu$ scattering from the bare deuterium nuclei are
employed~\cite{buba87,brac89}. At lower energies, where molecular
binding and electron-screening effects are significant, the differential
cross sections for $d\mu$ scattering from isolated D$_2$ molecules are
used~\cite{adam06}. The total ``molecular'' and ``nuclear'' cross
sections are smoothly sewed together, which was presented
in~Ref.~\cite{adam96}. The cross sections used take into account the
spin-flip of $d\mu$ atom and the rotational-vibrational excitations of
the target molecules. For the considered temperatures $T=40$--300~K, the
D$_2$ molecules are in the ground vibrational state $\nu=1$. The
population of the total spin states $F=1/2$ and~3/2 after the cascade is
assumed to be statistical.

\begin{figure}[htb]
   \centerline{\includegraphics[width=7cm]{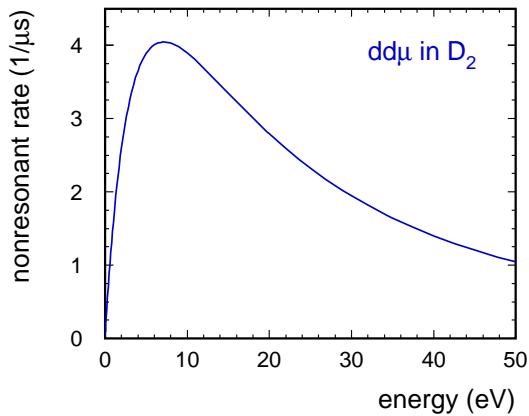}}
   \caption{The rate of nonresonant $dd\mu$ formation in $d\mu$ scattering
      from D$_2$ gas versus $d\mu$ energy.
      \label{fig:ddmnr}}
\end{figure}

The $d\mu$ atoms can form the muonic molecules $dd\mu$ in collision with
the D$_2$ molecules. Such a~process can have both the nonresonant or
resonant character~\cite{mens87}. The latter is possible due to the
presence of the loosely-bound state rotational-vibrational state $J=v=1$
of~$dd\mu$~\cite{vesm67}. The energy excess in nonresonant $dd\mu$
formation is taken away by the electron~\cite{faif89a}
\begin{equation*}
  d\mu+\mathrm{D}_2 \to [(dd\mu)de]^{+}+e\,.
\end{equation*}
The rate of this reaction is practically independent of the $d\mu$ atom
spin.  At $\varepsilon\lesssim{}100$~eV, only the s~and p~waves give
significant contributions to the formation rate.
In~Fig.~\ref{fig:ddmnr}, the calculated rate of nonresonant $dd\mu$
formation in $d\mu$ collision with a~free D$_2$ molecule is plotted
versus $d\mu$ energy~\cite{faif89a}. The formation rates in this paper
are given for the target density $\phi=1$ in the liquid hydrogen density
units~(LHD).
The nonresonant $dd\mu$ formation rate is significant
($\sim{}1\,\mu\mathrm{s}^{-1}$) even at a~few tens~eV, has maximum at
$\varepsilon\approx{}7$~eV, and falls rapidly when $\varepsilon\to{}0$.

\begin{figure}[htb]
   \centerline{\includegraphics[width=7.3cm]{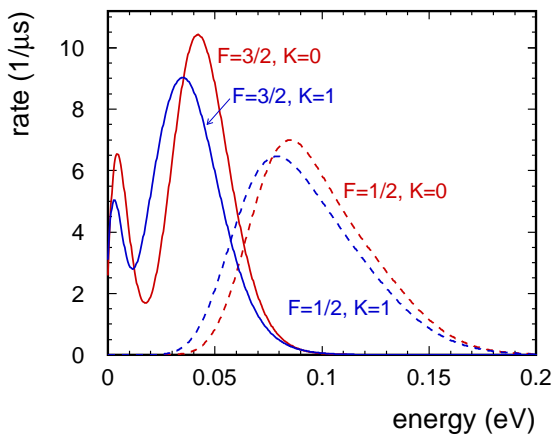}}
   \caption{The effective rate of resonant $dd\mu$ formation in $d\mu$ 
      scattering from 40-K ortho-D$_2$ ($K=0$) and para-D$_2$ ($K=1$) gas 
      versus $d\mu$ energy. The solid lines correspond to the upper spin 
      state $F=3/2$ of~$d\mu$, the dashed lines describe the $F=1/2$ state.
      \label{fig:ddmres_40k}}
\end{figure}
In the resonant $dd\mu$ formation, the energy excess is transferred to
rotational-vibrational excitations of the muonic-molecular complex
\begin{equation*}
  d\mu+\mathrm{D}_2 \to [(dd\mu)dee]^{*}.
\end{equation*}
The resonant rate strongly depends on the $d\mu$ energy and its total
spin state~$F$. In~Fig.~\ref{fig:ddmres_40k}, the effective resonant
$dd\mu$ rate (with the back decay of $dd\mu$ included) for the perfect
D$_2$-gas target at~40~K, calculated according to the method presented
in~Ref.~\cite{faif89}, is shown for the states $F=3/2$ and~1/2.
Comparing Figs.~\ref{fig:ddmnr} and~\ref{fig:ddmres_40k}, one can see
that the resonant formation dominates at lowest energies
($\varepsilon\lesssim{}0.1$~eV). The resonant rates calculated for the
rotational states $K=0$ and~1 of the target D$_2$ molecule display
significant differences. As a~result, the time spectra of $d$-$d$ fusion
products should expect appreciable ortho-para effect at low
temperatures.

The Monte Carlo simulations for the 40-K D$_2$ target have been
performed using the nonresonant $dd\mu$ formation rates shown
in~Fig.~\ref{fig:ddmnr}. The resonant $dd\mu$ formation has been
described using the absolute resonant formation rates and the back-decay
rates. Since the mean free path of~$d\mu$'s is much smaller than the
dimensions of the JINR target (the volume of~180~cm$^3$), the
simulations have been carried out for the infinite D$_2$ target. The
nitrogen contamination of~$10^{-6}$ have been taken into account. The
reliable theoretical distribution of $d\mu$ energy after the cascade
process is not calculated yet. Therefore, following Ref.~\cite{abbo97},
the initial $d\mu$ energy is assumed to be a~sum of the two Maxwell
distributions. The first Maxwell component corresponds to the
nonthermalized $d\mu$'s and the second one describes the thermalized
atoms. A~relative population of these two groups of $d\mu$'s is used as
a~free parameter in our simulations.

\begin{figure}[htb]
   \centerline{\includegraphics[width=7cm]{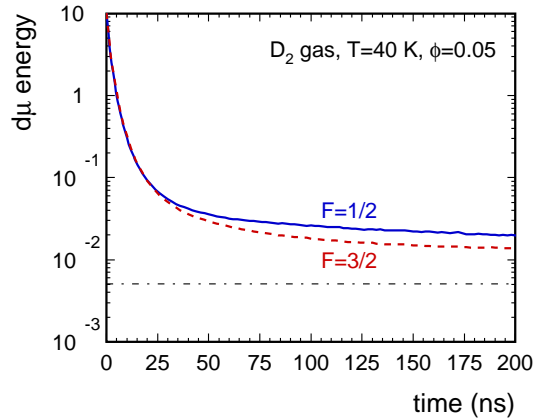}}
   \caption{Mean kinetic energy of $d\mu(F=1/2)$ and $d\mu(F=3/2)$
      atoms in D$_2$ gas ($T=40$~K, density $\phi=0.05$) versus time.
      The initial $d\mu$ energy distribution has the Maxwellian form
      with the mean energy of~10~eV.
      \label{fig:eav_200ns}}
\end{figure}

\begin{figure}[htb]
   \centerline{\includegraphics[width=8.8cm]{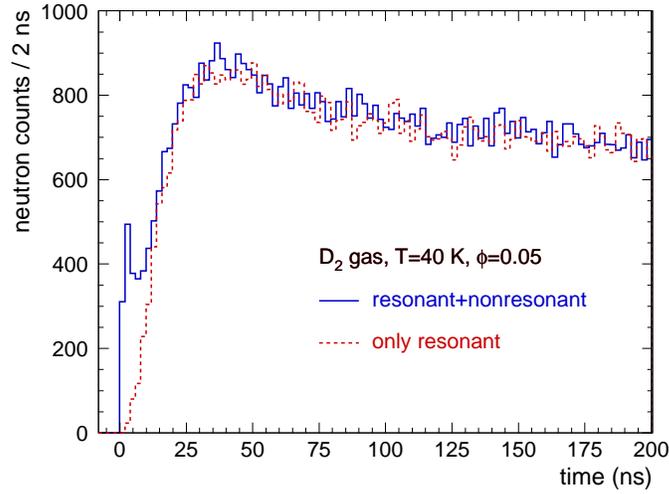}}
   \caption{Spectrum of neutrons from $d$-$d$ fusion in D$_2$ gas
            ($T=40$~K, density $\phi=0.05$) for short times.
            The contributions from all $dd\mu$ formation processes
            and from resonant process are shown separately.
      \label{fig:neut_res-nres}}
\end{figure}

\begin{figure}[htb]
   \centerline{\includegraphics[width=7.3cm]{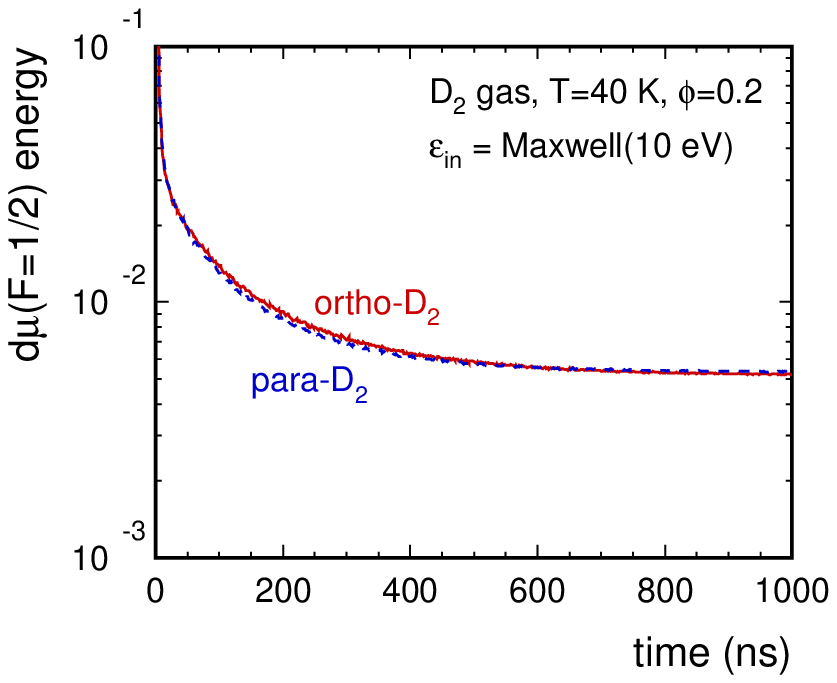}
               \hspace{5mm}
               \includegraphics[width=7.3cm]{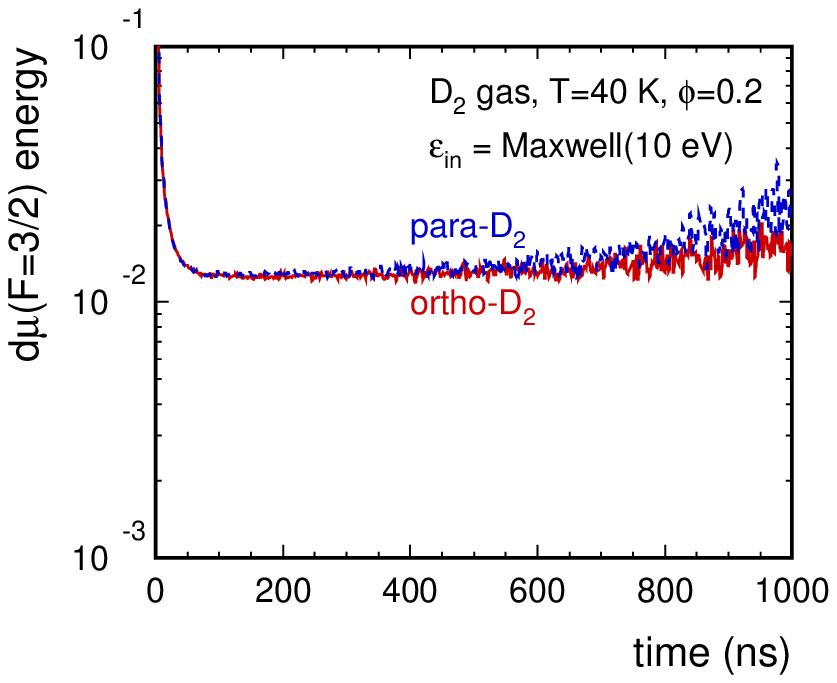}}
   \caption{Mean kinetic energy of $d\mu(F=1/2)$ and $d\mu(F=3/2)$
      atoms in ortho-D$_2$ and para-D$_2$ gas ($T=40$~K, density $\phi=0.2$)
      versus time. The initial $d\mu$ energy distribution has the Maxwellian
      form  with the mean energy of~10~eV.
      \label{fig:eav_ort-par}}
\end{figure}

\begin{figure}[htb]
   \centerline{\includegraphics[width=9cm]{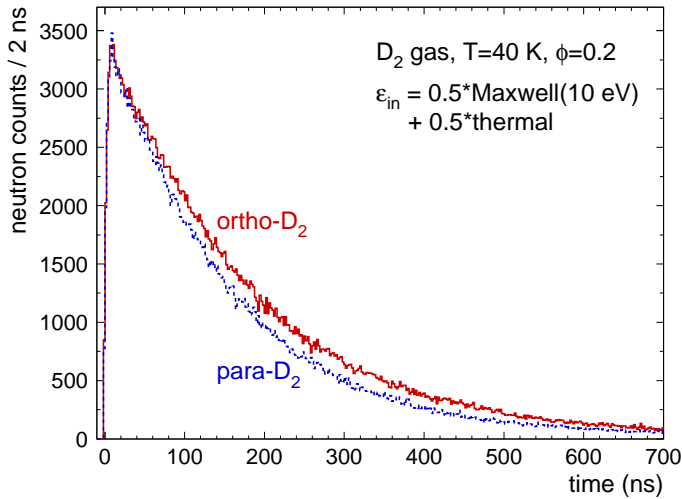}}
   \caption{Time spectrum of neutrons from $d$-$d$ fusion in ortho-D$_2$
     and para-D$_2$ gas at $T=40$~K and density $\phi=0.2$.
     \label{fig:neut_ort-par}}
\end{figure}

In Figs.~\ref{fig:eav_200ns}--\ref{fig:neut_ort-par} are shown some
results of our calculations for the 40-K D$_2$ gas.  The mean
energy~$\varepsilon_\text{avg}$ of $d\mu$'s in the both spin states is
shown in Fig.~\ref{fig:eav_200ns} for short times.
Only the nonthermalized Maxwell component
with~$\varepsilon_\text{avg}=10$~eV has been considered, in order to
study the $d\mu$ deceleration from high energies.  One sees that slowing
down to~0.05~eV is very fast --- it takes about~25~ns.  Then, this
process is much slower since $d\mu$'s can gain appreciable energy due to
the thermal motion of the target D$_2$ molecules. At this stage of
thermalization a~difference between the mean energies in the states
$F=1/2$ and~3/2 is apparent. This is caused by the downwards spin-flip
reaction
\begin{equation*}
  d\mu(F=3/2)+\mathrm{D}_2\to{}d\mu(F=1/2)+\mathrm{D}_2 \,,
\end{equation*}
in which the energy $\Delta{}E_\text{hfs}=0.0495$~eV (in the $d\mu+d$
center of mass) is released. The dash-dotted line shows the thermal
energy of~0.005~eV corresponding to~$T=40$~K.
The time spectra of neutrons from muon-catalyzed $d$-$d$ fusion for the
same initial conditions are shown in~Fig.~\ref{fig:neut_res-nres}. The
solid line represents the time spectrum with both the resonant and
nonresonant $dd\mu$ formation processes taken into account. The dashed
line has been obtained assuming only the presence of the resonant
formation. At short times ($\lesssim{}20$~ns) the fusion is mainly due
to the nonresonant $dd\mu$ formation since most $d\mu$'s are not slowed
down to the energies corresponding to the resonance peaks. This
phenomenon is more pronounced at lower target densities~\cite{Balin07}.

When the temperature of a~300-K D$_2$ gas is lowered, the even
rotational states deexcite to $K=0$ and the odd rotational states fall
to $K=1$, because of the symmetry of the D$_2$ wave function. As
a~result, the cooled deuterium target forms the statistical mixture
(2:1) of the ortho-D$_2$ and para-D$_2$ molecules~\cite{soue86}.  The
deceleration of $d\mu(F=3/2)$ and $d\mu(F=1/2)$ atoms in ortho-D$_2$ and
para-D$_2$ targets at $T=40~K$ and $\phi=0.2$ is illustrated
in~Fig.~\ref{fig:eav_ort-par}.  The ground-state $d\mu(F=1/2)$ atoms are
fully thermalized after about~1000~ns.  On the other hand, the mean
energy of the $d\mu(F=3/2)$ atoms never reaches the thermal value. The
initial population of these excited atoms disappears after a~few
hundreds~ns. For larger times, one sees only a~very small amount of the
$d\mu(F=3/2)$ atoms that are created via the back decay of the
muonic-molecular complex~$[(dd\mu)dee]$. In this process, they gain
appreciable kinetic energy.

Ortho-para effects in $\mu$CF in solid, liquid, and cold-gas D$_2$
targets with different concentrations of ortho-D$_2$ molecules have been
experimentally studied~\cite{toyo03,imao06}. The measured ortho-para
effects in solid deuterium targets (stronger resonant $dd\mu$ formation
in para-rich~D$_2$) are opposite to those predicted by
theory~\cite{adam01}, which requires further experimental and
theoretical investigations. On the other hand, theory and experiment
give the same sign of the ortho-para effect in the case of cold D$_2$
gases.

The calculated fusion-neutron time spectra for ortho-D$_2$ and
para-D$_2$ gas at $T=40$~K and $\varphi=0.2$ are shown
in~Fig.~\ref{fig:neut_ort-par}. The two-Maxwell
distribution~\cite{abbo97} of the initial $d\mu$ energy has been
assumed. After short time, the number of neutrons in ortho-D$_2$ case is
significantly greater than in the para-D$_2$ target. This difference
disappears at much larger times. Such a~bump in the ortho-rich D$_2$
target at similar conditions has been experimentally
observed~\cite{imao06}. Thus, the theory agrees with the experimental
findings, at least qualitatively.

\begin{figure}[htb]
   \centerline{\includegraphics[width=7.3cm]{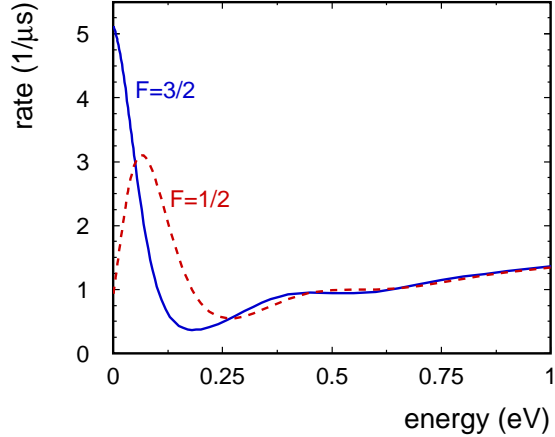}}
   \caption{The rate of resonant $dd\mu$ formation in $d\mu$ scattering
      from 300-K D$_2$ gas versus $d\mu$ energy. The solid lines correspond
      to the upper spin state $F=3/2$ of~$d\mu$, the dashed lines describe
      the $F=1/2$ state.
      \label{fig:ddmres_300k}}
\end{figure}

\begin{figure}[htb]
   \centerline{\includegraphics[width=7.5cm]{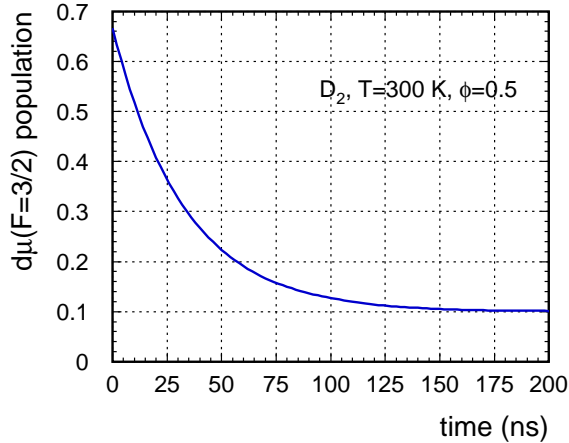}}
   \caption{Time evolution of the spin state $F=3/2$ fusion in D$_2$
     gas ($T=300$~K, density $\phi=0.5$).
     \label{fig:spinup_300k}}
\end{figure}

\begin{figure}[htb]
   \centerline{\includegraphics[width=8.8cm]{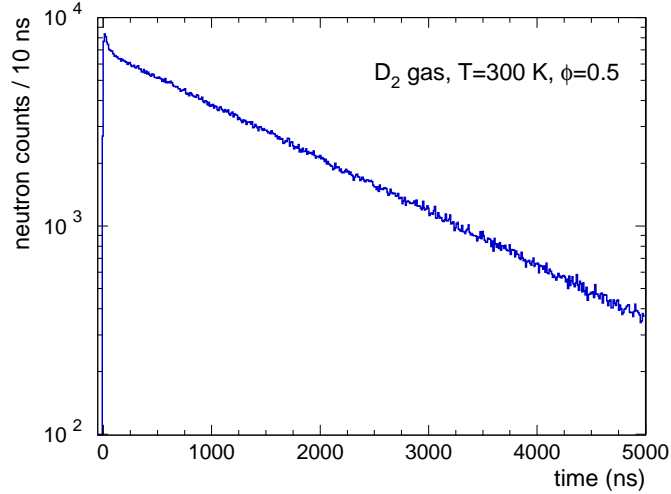}}
   \caption{Time spectrum of neutrons from $d$-$d$ fusion in D$_2$
     gas at $T=300$~K and at density $\phi=0.5$.
     \label{fig:neut_300k}}
\end{figure}

\begin{figure}[!htb]
   \centerline{\includegraphics[width=8.8cm]{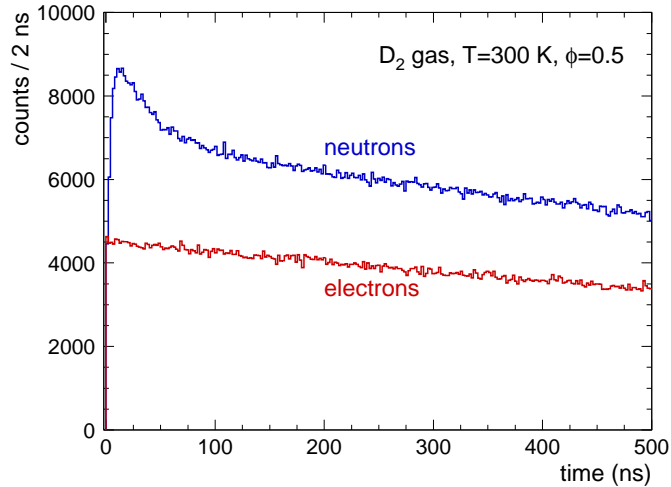}}
   \caption{Time spectra of fusion neutrons and of electrons from muon
     decay in D$_2$ gas at $T=300$~K and at density $\phi=0.5$.
     \label{fig:electron_300k}}
\end{figure}

In Figs.~\ref{fig:ddmres_300k}--\ref{fig:electron_300k} are presented
our theoretical results for the D$_2$ perfect-gas target at~300~K. The
effective resonant formation rates for $F=1/2$ and~3/2 are plotted
in~Fig.~\ref{fig:ddmres_300k}. These functions are more flat than those
in~Fig.~\ref{fig:ddmres_40k} since the thermal distribution of D$_2$
energies is much broader at~300~K and more initial rotational states are
significantly populated at this temperature.
Figure~\ref{fig:spinup_300k} shows the Monte Carlo time evolution of the
population of the $d\mu(F=3/2)$ atoms at~300~K and $\phi=0.5$. The
initial statistical value of the upper-spin population is equal to~2/3.
After about 200~ns, the thermal equilibrium between the lower and upper
spin states is reached.  The theoretical downwards spin-flip rate has
been scaled by the constant factor of~0.6, determined by comparison with
the experimental results~\cite{voro01}.
The thermalization of $d\mu$ atoms at such a~density is very fast --- it
takes only a~few~ns. Therefore, the initial $d\mu$-energy distribution
does not affect the fusion-product spectra at significantly larger
times.

The calculated neutron time spectrum for this D$_2$ target is plotted
in~Fig.~\ref{fig:neut_300k}, for very large time interval. Apart from
the prompt peak, the spectrum can be well described by the steady-state
kinetics.
The Monte Carlo spectra of fusion neutrons and electrons from muon
decays are shown for short times in~Fig.~\ref{fig:electron_300k}. The
electron spectrum is flat since the muon decay does not depend on the
$d\mu$ atom energy.

In conclusion, the theoretical results of Monte Carlo simulations of the
$\mu$CF cycle processes in gaseous D$_2$ targets, corresponding to the
JINR target conditions, have been obtained. The theoretical resonant and
nonresonant $dd\mu$ formation rates and the differential cross sections
for $d\mu$ scattering from D$_2$ molecules established the input for our
calculations. In particular, the thermalization of $d\mu$ atoms for
various targets and the role of nonresonant $dd\mu$ formation in
short-time neutron spectra have been investigated. The nonresonant
formation can be directly observed at very short times in
low-temperature targets at densities $\phi\lesssim{}0.05$. Also, the
$d\mu$-thermalization time is important at relatively low target
densities. The calculated neutron time spectra for ortho-D$_2$ and
para-D$_2$ at 40~K and~$\phi=0.2$ display the enhanced neutron emission
in the ortho-rich targets at intermediate times, which agrees
qualitatively with the recent experimental data. The simulation for the
high density ($\phi=0.5$) D$_2$ target at~300~K shows that the neutron
time spectrum can be well described by the steady-state kinetics, apart
from very short times.

\vspace{20pt}
This work was supported by the INTAS grant Nr~05-1000008-7953.


\end{document}